\documentclass[12pt,aps,
amsmath,
]{article}
\usepackage{bm}
\usepackage{graphicx}

\usepackage{amssymb}

\begin{document}

\title{Statistical Properties of  the T-exponential of
Isotropically Distributed Random Matrices.}

\author{A.S. Il'yn$^{1,2}$\footnote{E-mails: asil72@mail.ru, sirota@lpi.ru, zybin@lpi.ru},
V.A. Sirota$^{1}$, K.P. Zybin$^{1,3}$ }
\date
{\small $^1$ P.N.Lebedev Physical Institute of RAS, 119991,
Leninskij pr.53, Moscow, Russian Federation \\ $^2$ National
Research Nuclear University MEPhI, 115409, Kashirskoe shosse, 31
Moscow, Russian
Federation  \\
 $^3$ National Research University Higher
School of Economics, 101000, Myasnitskaya 20, Moscow, Russian
Federation}

\maketitle

\begin{abstract}
A functional method for calculating averages of the time-ordered
exponential of a  continuous isotropic random $N\times N$ matrix
process is presented. The process is not assumed to be Gaussian. In
particular, the Lyapunov exponents and higher correlation functions
of the T-exponent are derived from the statistical properties of the
process.

The approach may be of use in a wide range of physical problems. For
example, in theory of turbulence the account of non-gaussian
statistics is very important since  the non-Gaussian behavior is
responsible for the time asymmetry of the energy flow.

{ \small Keywords: {Lyapunov exponents \and random matrices \and
T-exponential \and  functional integral \and stochastic equations
\and turbulence}
 }
\end{abstract}


\section{ Introduction}

Sets of  linear stochastic differential equations appear in
different physical problems related to quantum mechanics and field
theory, turbulence, low temperature physics etc. Their formal
solution is given by the time-ordered exponential, but calculation
of the statistical moments is still a challenge.

The long-time evolution of the time-ordered product of arbitrary
(not Gaussian) random matrices in the discrete case was investigated
in \cite{Let1}, \cite{Let2}, and \cite{Let3}. The existence of the
Lyapunov spectrum was proved, though no recipe to calculate the
Lyapunov exponents or other averages was given.
A functional integration method to calculate the averages in the
case of  Gaussian $\delta$-correlated processes was introduced in
\cite{GambaKol} and \cite{Gamba}.

However, in many applications  one needs to calculate the
T-exponentials of non-Gaussian processes. These are, e.g., field
theories with interactions. Also, all the processes that produce
non-zero correlators of odd orders, or those in which the Lyapunov
spectrum is not even, are knowingly non-Gaussian. This is just what
occurs in the theory of magnetic dynamo (see \cite{zeld}) and in the
theory of turbulence: the third-order correlator is related to the
energy dissipation rate (\cite{kolm,Frisch}), and the asymmetry of
the Lyapunov spectrum is necessary to provide the time anisotropy of
a turbulent flow (\cite{ZSpre,IZ}). Besides, the velocity gradient
tensor, which  is an important  object in the problems related to
passive scalar (\cite{BalkFoux,FalkGawVerg}) and in derivation of
scaling exponents (\cite{ZSpre,ufnZS}), is proved to have
non-Gaussian distribution (\cite{nonGauss}). In this paper we
propose a functional integration method for calculation the averages
of continuous products of $N\times N$ random matrices $A(t)$. The
probability distribution of the matrices is assumed to be isotropic,
Gaussianity is not required. The method simplifies the functional
integrals significantly, in the case of $\delta$-correlation it
allows to express all the statistical characteristics of the
T-exponential in terms of the moments of $A(t)$. The main idea of
the approach is to decompose the matrix product into the rotational
and deformational components, and then change the variables.
Although the change of variables is non-local (relative to $t$),  it
allows to minimize the non-locality and exclude the time ordering
from the path integral. One more important point is that  we use a
cumulant function to analyze the $\delta$-processes. This allows to
avoid renormalizations, and to get a simple expressions for all the
averages.

\section{ T-exponential of a random process}

Let $A(t)$, $0\le t \le T$ be a random process taking on a value of
$N \times N$ real matrices. Its statistics is defined by the measure
$$
DA\,P\left[ A \right] \equiv \prod\limits_{0 \le t \le T}
{\,\prod\limits_{k,p = 1}^N {d{A_{kp}}\,P\left[ A \right]} ,\,\,\,}
$$
 where $P[A]$  is the probability density functional.

Consider the random matrices $Q(t)$  that satisfy the equation
\begin{equation} \label{Qequation}
{\partial _t}Q = QA,\,\,Q\left( 0 \right) = \hat 1
\end{equation}
The formal solution to the equation can be written in terms of the
anti-chronological exponential (\cite{antiT}):
\begin{equation} \label{Q-Texp}
Q\left( t \right) = \mathop T\limits^ +  \exp \left(
{\int\limits_0^t {A\left( \tau \right)d\tau } } \right) =
\sum\limits_n {\frac{1}{{n!}} \int\limits_0^t {d{\tau _1}...d{\tau
_n}\mathop T\limits^ +
 \left( {A\left( {{\tau _1}} \right)...A\left( {{\tau _n}} \right)} \right)} }
 \end{equation}
where
\[\mathop T\limits^ +  \left( {A\left( {{\tau _1}} \right) \dots A\left( {{\tau _n}} \right)}
\right) = A\left( {{\tau _{{i_1}}}} \right)...A\left( {{\tau
_{{i_n}}}} \right), \ \quad {\tau _{{i_1}}} \le ... \le {\tau
_{{i_n}}}\] is the antichronological product operator. The
alternative way to describe the solution is the Volterra
multiplicative integral (\cite{Gantmacher-Volterra}):
\begin{equation} \label{Volterra}
Q\left( t \right) = \prod\limits_{\tau  = 0}^t {\left( {1 + A\left(
\tau \right)d\tau } \right)}
\end{equation}
 Thus, the T-exponent is equivalent to the infinite matrix
product.

We are interested in the averages
\begin{equation} \label{averageA}
\left\langle {F\left[ Q \right]} \right\rangle  = \int
{DA\,\,P\left[ A \right]F\left[ Q \right]} \ ,
\end{equation}
 where $F[Q]$ is some
functional. These expressions contain  the T-exponent which is not
an easy object to  deal with. Below we simplify them and reduce, in
the case of isotropically distributed processes,  to the path
integrals of some exponents without any time-ordering.

\section{Change of variables}

We make  the Iwasawa decomposition  of the matrix $Q$:
\begin{equation}    \label{Iwasawa}
Q = z\,d\,R
\end{equation}
where $z$ is an upper triangular matrix with diagonal elements equal
to 1,  $d$ is a diagonal matrix,  $R$  is an orthogonal matrix:
$R_{ji} R_{jk}=\delta_{ik}$.

We recall that $Q$ as well as $z,d,R$ is a function of time. From
(\ref{Qequation}) it follows $A = {Q^{ - 1}}{\partial _t}Q$; thus,
Eq. (\ref{Iwasawa}) produces the decomposition of $A$:
\begin{equation} \label{decA}
A = {R^T}\, X \,R  \ , \qquad X=\rho  + \zeta + \theta
\end{equation}
where
\begin{equation} \label{rho-and-x}
\rho  = {d^{ - 1}}{\partial _t}d \  , \qquad \zeta = {d^{ - 1}}{z^{
- 1}}\left( {{\partial _t}z} \right)d
\end{equation}
\begin{equation} \label{thetaR}
\theta  = \left( {{\partial _t}R} \right){R^T}
\end{equation}
We see that $\rho$ is diagonal, $\zeta$ is an upper triangular
matrix with zeroes in the main diagonal, and $\theta$ is
antisymmetric:
$$
\rho=diag(\rho_1,\dots,\rho_N)  \ , \qquad \theta_{ij}=-\theta_{ji}
\ , \qquad \zeta_{ij}=0 \ \ \mbox{if} \ \ i\ge j
$$

Now we consider $X(t)$ as independent functional variables; then
(\ref{decA}) should be understood as
\begin{equation} \label{change}
A = {R^T}\left[ X \right]\,\,X\,\,R\left[ X \right]
\end{equation}
 Where $R[X]$ is
determined by (\ref{thetaR}):
\begin{equation} \label{R-Texp}
R\left( t \right) = T\exp \left( {\int\limits_0^t {\theta \left(
\tau  \right)d\tau } } \right)
\end{equation}
Note that $R$ depends only on the $\theta$-component of $X$; this
will simplify further calculations significantly. This also makes it
possible to separate the rotational part of $Q$ from deformation.

The averages (\ref{averageA}) can  be rewritten as
\begin{equation} \label{averageX}
\left\langle {F\left[ Q \right]} \right\rangle  = \int {DX\, J[X] \;
P\left[ R^T[X] \, X \, R[X]  \right] \;   F\left[ Q[X] \right]}
\end{equation}
The Jacobian $J[X]$ will be calculated in the next section.

The advantage of  this change of variables is in dealing with a
simple 'rotational' T-exponent  (\ref{R-Texp}), instead of the
complicated T-exponent (\ref{Q-Texp}).  As we will see below,  this
one does also vanish in the isotropic case.

\section{The  Jacobian}

In this section, we calculate the functional Jacobian
\[J = Det\left( {\frac{{\delta {A_{ij}}\left( t \right)}}{{\delta {X_{kp}}\left( {t'} \right)}}} \right)\]
First, to simplify the notations, we introduce the multiindices
(hereafter denoted by Greek letters):
\[ \alpha  \equiv  \left( {i,j} \right)\  ; \qquad A_{\alpha} \equiv A_{ij} \ , \qquad
 X_{\beta} \equiv X_{kp} \ ,  \qquad  \alpha , \beta , \dots = 1..{N^2}\]
The transformation (\ref{change}) can  be presented as
\begin{equation} \label{multiA}
{A_\alpha } = {\Re _{\alpha \beta }}\left[ X \right]{X_\beta } \ ,
\end{equation}
where $\Re _{\alpha \beta } [X]$ is the $N^2\times N^2$ matrix that
 satisfies the relation:
\begin{equation} \label{Rmulti}
{\Re _{\left( {ij} \right),\left( {kp} \right)}} = {R_{ki}}{R_{pj}}
\end{equation}
(This corresponds to multiplying $X$ by $\Re$  from right and by
$R^T$ from left). The functional derivative of (\ref{multiA}) is
 equal to
\begin{equation} \label{Jacmatr}
\frac{{\delta {A_\alpha }\left( t \right)}}{{\delta {X_\beta }\left(
{t'} \right)}} = {\Re _{\alpha \beta }}\left( t \right)\delta \left(
{t - t'} \right) + \frac{{\delta {\Re _{\alpha \gamma }}\left( t
\right)}}{{\delta {X_\beta }\left( {t'} \right)}}{X_\gamma }\left( t
\right)
\end{equation}
Now, as it follows from (\ref{Rmulti}) and (\ref{R-Texp}),  $\Re(t)$
is determined completely by the  function $\theta(\tau)$ at $\tau
\le t$. Thus, $\Re_{\alpha \beta} (t)$ does not depend on
$X_{\gamma}(t')$ if $t<t'$.

According to (\ref{Jacmatr}), the Jacobian is a block-triangular
matrix:
\[\left( {\frac{{\delta {A_\alpha }\left( t \right)}}{{\delta {X_\beta }\left( {t'} \right)}}} \right) = 0,\,\, \quad  t < t'\]
Consequently  , its determinant is a continuous product of the
determinants of the diagonal ($t=t'$)  $N^2\times N^2$ blocks:
\begin{equation} \label{Jprod}
J  \equiv Det\left( {\frac{{\delta {A_\alpha }\left( t
\right)}}{{\delta {X_\beta }\left( {t'} \right)}}} \right) =
\prod\limits_{0 \le t \le T} \tilde J \left( t \right)   \  , \qquad
\tilde J ( t ) = \det \left( \frac{\delta {A_\alpha } (t)}{\delta
{X_\beta } (t)} \right)
\end{equation}
\vspace{0.7cm}

Now, let us calculate $\tilde J ( t )$. From (\ref{Jacmatr}) it
follows
\[\tilde J\left( t \right) = \det \left( {{\Re _{\alpha \beta }}\left( t \right)\delta \left( 0 \right) + \frac{{\delta {\Re _{\alpha \gamma }}\left( t \right)}}{{\delta {X_\beta }\left( t \right)}}{X_\gamma }\left( t \right)} \right)\]
Note that $\Re_{\alpha \beta }$ is an orthogonal matrix, since
\[\,{\Re _{\mu \alpha }}{\Re _{\mu \beta }} \equiv {\Re _{\left( {ij} \right),
\left( {kp} \right)}}{\Re _{\left( {ij} \right),\left( {mn}
\right)}}
 = {R_{ki}}{R_{pj}}{R_{mi}}{R_{nj}} = {\delta _{km}}{\delta _{pn}}  =  {\delta _{\left(
 {kp} \right),\left( {mn} \right)}} = {\delta _{\alpha \beta }}\]
So, $\det  \Re =1$, and one can multiply the bracketed expression by
$\Re ^T$. Then, assuming $\delta \left( 0 \right) = {\left( {dt}
\right)^{ - 1}}$ and  omitting the insufficient normalization
multipliers, we obtain
$$
\tilde J\left( t \right) = \det \left( {{\delta _{\alpha \beta }} +
{\Re _{\mu \alpha }}\frac{{\delta {\Re _{\mu \gamma }}}}{{\delta
{X_\beta }}}{X_\gamma }dt} \right) =\exp  \left(  \mbox{tr}   \left(
{{\Re _{\mu \alpha }}\frac{{\delta {\Re _{\mu \gamma }}}}{{\delta
{X_\beta }}}{X_\gamma }}    \right) dt \right)
$$
From (\ref{Jprod})  it  follows
\begin{equation} \label{JG}
J = \exp \left( {\int\limits_0^T {G\,dt} } \right) \ , \qquad G =
\Re _{\mu \alpha }  \frac{\delta {\Re _{\mu \gamma }}}{\delta
{X_\alpha }}{X_\gamma }
\end{equation}
\vspace{0.5cm}

To calculate $G$, we return from multiindices to usual matrix
notations:
\begin{equation} \label{G}
G = {R_{ki}}{R_{pj}}\frac{{\delta \left( {{R_{ni}}{R_{mj}}}
\right)}}{{\delta {X_{kp}}}}{X_{nm}}
\end{equation}

Now we need to calculate the variation derivative  $\frac{\delta
R_{nm} (t)}{\delta X_{kp}(t)} $ at coinciding time. The  $R(t)$
dependence of $X(t)$  is determined by (\ref{decA}), (\ref{R-Texp}).
Thus, of all the $N^2$ components of the matrix $X$, $R$ depends on
$N(N-1)/2$ independent components of $\theta_{ij}$ only.

Since $X_{ij}$ coincides with $\theta_{ij}$ as $i>j$, we get
$$
\frac{\delta \theta_{ij} (t)}{\delta X_{kp}(t)}  = \left\{
\begin{array}{lcl}  \delta_{ik} \delta_{jp} & if  & k>p \\  0 & if &
k \le p \end{array} \right.
$$
Hence,  $\frac{\delta R_{ij} (t)}{\delta X_{kp}(t)} =0 $ if $k\le
p$.

For the rest, we use the Volterra presentation (\ref{Volterra}) of
the T-exponent (\ref{R-Texp}):
\begin{eqnarray} \nonumber
\left. \frac{\delta R_{ij} (t)}{\delta \theta_{kp}(t)} \right|
_{k>p} =&& \frac{\delta}{\delta \theta_{kp}(t)}  \left(   \prod
\limits_{\tau=t}^{0} \left(1+\theta(\tau) d\tau \right) \right)
_{ij} = \\
\nonumber && \int \limits _0 ^t   \left(    \prod
\limits_{\tau=t}^{\tau'} \left( 1+\theta (\tau) d\tau \right)
\right)_{im} \;  \frac{\delta \theta_{mn}(\tau')}{\delta
\theta_{kp}(t)} d\tau' \;
 \left(    \prod \limits_{\tau=\tau'}^{0} \left(
1+\theta (\tau) d\tau \right) \right)_{nj}
\end{eqnarray}
Further,
$$
\frac{\delta \theta_{mn}(\tau')}{\delta \theta_{kp}(t)} =
\delta(\tau'-t) \left(  \delta_{mk} \delta_{np} - \delta_{mp}
\delta_{nk}  \right) \ ,
$$
and we make use of the time-ordering and note that the derivative
can only make a non-zero contribution when acting on the first
($\tau'=t$) multiplier.
 Then the rest of the multipliers make $R$ again, and
 $\int _0 ^t \delta(t-\tau') d\tau' = \frac 12$ because $t$ is the boundary
of the integral. (Effectively, only one half of the delta-function
is integrated.) The integral becomes more compact:
$$
\left. \frac{\delta R_{ij} (t)}{\delta \theta_{kp}(t)} \right|
_{k>p} = \frac 12 \left( \delta_{ik} \delta_{mp} - \delta_{ip}
\delta_{mk} \right) \left( \prod \limits_{\tau=t}^{0} \left(
1+\theta (\tau) d\tau \right) \right)_{mj} = \frac 12 \left(
\delta_{ik} \delta_{mp} - \delta_{ip} \delta_{mk} \right) R_{mj}
$$

Thus,
$$
\frac{\delta R_{ij}(t)}{\delta X_{kp}(t)} =  \left\{
\begin{array}{ll} \frac 12 \left( \delta_{ik} \delta_{mp} - \delta_{ip}
\delta_{mk}  \right) R_{mj}  \ ,& \quad k > p \\ 0 \ ,& \quad k\le p
\end{array} \right.
$$
As we substitute  this into (\ref{G}) and recall that $R^T R=I$, all
the matrices $R$ are cancelled, and we get
\[G = \sum\limits_{k > p} \frac{\left( X_{kk} - X_{pp} \right)}{2}
 =  tr \left(  \eta_0 X \right) \]
where
\begin{equation} \label{eta0}
 \left( \eta_0 \right)_{kp} = \,\frac{2k - 1 - N}{2} \delta _{kp}
\end{equation}
From (\ref{JG}) we eventually have
\begin{equation} \label{Jac}
J\left[ X \right] = \exp \left( \int\limits_0^T \mbox{tr} \left(
\eta_0 X (t)  \right) dt  \right)
\end{equation}

\section{Isotropic processes}

Now we  express the probability functional in the form
\[P\left[ A \right] = \exp \left( { - \int\limits_0^T
{L\left( {A,\,\,{\partial _t}A,\,\,\partial _t^2A,\,\,\dots}
\right)dt} } \right) \ , \] where $L$ is the Lagrangian of the
process.

The change  of variables (\ref{change}) transforms $P[A]\, DA$ to
$P_X[X]\, DX$ where
\[{P_X}\left[ X \right] = P\left[ {{R^T}XR} \right]J = \exp \left(
{ - \int\limits_0^T {L_X dt}  + \int\limits_0^T {tr\left( { \eta_0
X} \right)dt} } \right)\] Here
\[{L_X}= L\left( {{R^T}XR,\,\,{\partial _t}\left( {{R^T}XR} \right),\,\,\partial _t^2\left(
{{R^T}XR} \right),\,\,\dots} \right)\] contains generally not only
derivatives but also integrals of $X(t)$, since $R[X]$ contains the
T-exponent (\ref{R-Texp}). Hereafter, we restrict our consideration
with the isotropic processes:
\begin{equation} \label{isotr}
P\left[ {{O^T}AO} \right] = P\left[ A \right] \ \  \forall O \in
SO(N)
\end{equation}
(This means that statistical properties of the process would not
change under the global rotation of the reference frame.) We will
show that for such processes, $L_X$ does not contain the time
integrals.

The condition (\ref{isotr}) implies that the matrix $A$ and its time
derivatives can  contribute to $L$ only in scalar combinations like
\begin{equation} \label{examp}
 \mbox{tr}  \left(   A^{b_0}....\left( \partial _t^{a_1} A\right)^{b_1}
 \left( \partial _t^{a_2} A^T  \right)^{b_2} \dots  \right)
\end{equation}
But from  (\ref{thetaR}) it follows
$$
\partial_t R = \theta R \ , \quad   \partial^2_t R =\left(  \partial_t \theta +\theta^2 \right)
 R \ , \quad  \partial_t R^T =  R^T \theta^T \ , \dots ;
$$
accordingly,
$$
\partial_t \left( R^T XR \right) = R^T \left( \theta^T X + \partial_t X + X \theta \right)
R \  , \quad
\partial_t^a (R^T X R) = R^T \left( \dots \right) R
$$
Hence, as we substitute $R^T X R$ for $A$ in (\ref{examp}),  all the
$R^T$ are multiplied by $R$ and vanish. As a result, $L_X$  contains
scalar combinations of $X$ and $\theta$ and their derivatives, and
does not contain $R$.  According to (\ref{decA}), the matrix
$\theta$ is itself a function of $X$, so
\[{L_X}\left[ X \right] = {L_X}\left( {X,\,\,{\partial _t}X,\,\,\partial _t^2 X\,,\dots}
\right)\] Thus, the new Lagrangian $L_X$ is a differential function
of the matrix $X$.

\section{Isotropic $\delta$-processes.}

There is an important particular case of absence of correlation
between the values of $A$ at different time moments:
\begin{equation} \label{deltapr}
P\left[ A \right] = \exp \left( { - \int\limits_0^T {L\left( A
\right)dt} } \ , \right)
\end{equation}
 the Lagrangian $L$  being a
function of $A$ only, not of its derivatives. We also demand that
$A$ is isotropic (\ref{isotr}). Then $A$ can contribute to $L$ only
as a part of invariant combinations:
\[L ( A ) = L\left( {tr\,A,\,\,tr\,{A^2},\,\,tr\,A{A^T},\,tr\,{A^3}....} \right)\]
The change of variables  $A \mapsto X$  results in the substitution
$X$ for $A$, since all the $R$ and $R^T$ vanish. Thus,
\[{L_X}\left[ X \right] = \left. L ( A ) \right|_{A=X}\]
and
\begin{equation} \label{Pdelta}
{P_X}\left[ X \right] = \exp \left( { - \int\limits_0^T {L\left( X
\right)dt}  + \int\limits_0^T {tr\left( { \eta_0  X} \right)dt} }
\right)
\end{equation}

To get the correlation functions of $X$, one now just has to
calculate the functional integral (\ref{averageX}):
\begin{equation} \label{Xav}
\left\langle {{X_{ij}}\left( {{t_1}} \right)...{X_{kp}}\left(
{{t_n}} \right)} \right\rangle  = N'\int DX\,{X_{ij}}\left( {{t_1}}
\right)...{X_{kp}}\left( t_n \right) %
{P_X}\left[ X \right]
\end{equation}
where
\[N' = {\left( {\int {DX\,\exp \left( { - \int\limits_0^T {L\left( X \right)dt +
\int\limits_0^T {tr\left( {\eta_0 X} \right)dt} } } \right)} }
\right)^{ - 1}}\] Since $X$ is an integration variable, it can be
changed to $A$ in the right-hand side of the expression. Note that
the integral does not contain T-exponents.

\vspace{0.7cm}

To simplify the calculations, one  introduces the characteristic
functional of a random process:
\begin{equation}  \label{Z}
Z\left[ {\eta \left( t \right)} \right] = \left\langle {\exp \left(
{\int\limits_0^T {tr\left( {\eta \left( t \right)A\left( t \right)}
\right)dt} }
 \right)} \right\rangle  = \int {DA\,P[A]\exp \left( { \int\limits_0^T {tr\left(
 \eta A \right) dt} }  \right)}
 \end{equation}
instead of $P[A]$. The averages can then be expressed as
\begin{equation} \label{AZ}
\langle A_{ij} \left( t_1 \right) \dots A_{kp} \left( t_n \right)
\rangle  = \left. \frac{1}{Z}\frac{\delta }{\delta \eta _{ij}\left(
t_1 \right)} \dots \frac{\delta }{\delta \eta _{kp}\left( t_n
\right)}Z \right|_{\eta(t)=0}
\end{equation}
To determine the statistics of $Q$ (\ref{Q-Texp}), (\ref{averageA}),
we have to calculate different correlators of $X$. From (\ref{Xav})
it then follows that for $\delta$-processes the correlation function
of $X$ can also be written in terms of the same $Z[\eta]$:
\begin{equation} \label{XZ}
\left\langle {{X_{ij}}\left( {{t_1}} \right)...{X_{kp}}\left(
{{t_n}} \right)}
 \right\rangle  = \left.
 \frac{1}{Z}\frac{\delta }{{\delta {\eta _{ij}}\left( {{t_1}}
 \right)}} \dots
 \frac{\delta }{{\delta {\eta _{kp}}\left( {{t_n}} \right)}}Z  \right|_{\eta(t)=\eta_0}
\end{equation}
The only difference from (\ref{AZ}) is that the expression is
calculated at the point $\eta (t)= \eta_0$ instead of $\eta (t)=0$.

It is convenient to introduce the generating functional  for
connected correlation functions defined by
\begin{equation} \label{W}
Z[\eta] = e^{W[\eta]}
\end{equation}
The normalization requires $W[\eta(t)=0]=0$. Since $Z[\eta]$ is a
Fourier transform of $P[A]$, the isotropy (\ref{isotr})  of $P[A]$
 leads to isotropy of $Z[\eta]$ and $W[\eta]$:
$$ W\left[ {{O^T} \eta O} \right] = W\left[ \eta \right] \ \  \forall O \in
SO(N)
$$
Furthermore, if $P[A]$ is  a $\delta$-process (\ref{deltapr}), $Z$
is also a continuous product of independent multipliers, and
\begin{equation} \label{wdef}
W\left[ {\eta \left( t \right)} \right] = \int\limits_0^T {w\left(
{\eta \left( t \right)} \right)dt}
\end{equation}
The function
$w(\eta)$ is called  a cumulant function (see, e.g.,
\cite{Klyackin}).
 Via this function, one can
calculate any of the correlators by consecutive differentiation:
\begin{equation} \label{Aw}
{\left\langle {{A_{ij}}\left( {{t_1}} \right)...{A_{kp}}\left(
{{t_n}} \right)} \right\rangle } = \left. \frac{\delta }{{\delta
{\eta _{ij}} \left( {{t_1}} \right)}}...\frac{\delta }{{\delta {\eta
_{kp}}\left( {{t_n}} \right)}}
 \exp \left({\int \limits_0^T w(\eta) dt} \right) \right| _{\eta(t)=0}
 \end{equation}
The first term of each correlator,
$$ \left\langle {{A_{ij}}\left( {{t_1}} \right)...{A_{kp}}\left(
{{t_n}} \right)} \right\rangle_c = \frac{\partial^n w}{\partial
{\eta _{ij}} ...\partial  {\eta _{kp}} }(0) \delta(t_2-t_1) \delta
(t_3-t_1) \dots \delta(t_n-t_1) \ ,
$$
 is called the connected correlation function and
 corresponds to the connected
diagram (as $t_1$,\dots, $t_n$ are represented by $n$ points, and
$\delta$-functions make connections between them). The next terms
contain the products of lower-order connected correlation functions
(and hence, smaller sets of $\delta$-functions) and correspond to
non-connected diagrams.

By analogy to (\ref{XZ}),  all the correlators of the $X$-variables
can easily be obtained from the same expressions as these for
$A$-variables by changing the point where the derivatives are taken.
For the connected correlation functions we get:
\begin{equation} \label{Xw}
\left\langle {{X_{ij}}\left( {{t_1}} \right)...{X_{kp}}\left(
{{t_n}} \right)} \right\rangle _c =   \frac{\partial ^n w}{\partial
\eta _{ij} ...\partial \eta _{kp}}
  (\eta_0)  \delta(t_2-t_1) \delta (t_3-t_1) \dots
\delta(t_n-t_1)
 \end{equation}

\section{The Lyapunov spectrum}

In this section we use the method described above to calculate the
averages of the matrix elements $X_{ij}$ in the case of isotropic
$\delta$-processes. From (\ref{Xw}) it follows
\begin{equation}  \label{lamb1}
\left\langle {{X_{sq}}} \right\rangle  =  \frac{\partial }{{\partial
{\eta _{sq}}}}w\left( { \eta_0 } \right)
\end{equation}

First, we show that the non-diagonal matrix elements are equal to
zero. Actually, since $w(\eta)$ is isotropic, it depends on a
combination of traces:
\begin{equation} \label{wisotr}
w\left( \eta  \right) = w\left( {tr\,\eta ,\,\,tr\,{\eta
^2},\,\,tr\,\eta {\eta ^T},\,tr\,{\eta ^3}....} \right)
\end{equation}
 The derivative of each trace is
$$
\frac{\partial }{\partial {\eta _{sq}}} tr \left( \eta \dots \eta^T
\dots\right) =\sum\delta_{is} \delta_{jq}  \left( \eta \dots \eta^T
\dots \right)_{ji}
$$
Since 
the derivative is taken at $ \eta _{kp} =  \left(\eta_0 \right)_{kp}
\sim \delta _{kp}$ (see (\ref{eta0})) , we get
$\delta_{is}\delta_{jq} \delta_{ij}$ in each term, and non-diagonal
elements vanish:
$$
\left\langle {{X_{sq}}} \right\rangle  = \frac{\partial }{{\partial
{\eta _{sq}}}}w\left( { \eta_0 } \right) = 0 \ , \qquad s \ne q
$$

The averages of the diagonal components,
\begin{equation} \label{lambdadef}
 {\lambda_s} = \left\langle {{X_{ss}}} \right\rangle  = \left\langle {{\rho
_s}} \right\rangle  \qquad (no \  summation)
\end{equation}
are called the  Lyapunov exponents. The set of $\lambda_s$ is an
important statistical characteristic of a process.  It is used in
many physical applications (e.g., to describe the separation of
trajectories in a turbulent  flow).

Before we proceed to these calculations, we make one useful deduction:  \\
For the values $\lambda_s$, as well as other diagonal averages, the
function $w(\eta)$ in (\ref{Xw}), (\ref{lamb1}) can be replaced by
the diagonal cumulant function $w_d$ where all the non-zero elements
are set equal to zeros:
\begin{equation} \label{wdiag}
{w_d}\left( {{\eta _{11}},...,{\eta _{NN}}} \right) = {\left.
{w\left( \eta  \right)} \right|_{{\eta _{kp}} = 0,k \ne p}}
\end{equation}
This is caused by the diagonality of $\eta_0$ so in (\ref{Xav}) one
can make an
integration over the non-diagonal elements to get the 'diagonal' probability function %
%
%
$$
{P_d}\left[ A_{11},\dots, A_{NN} \right] = \int {\prod\limits_{k \ne
p} {D{A_{kp}}\exp \left( { - \int\limits_0^T {L\left( A \right)dt} }
\right)} }
$$
The cumulant function (\ref{wdiag}) corresponds to this PDF.

\subsection{Gaussian process}

Let the probability distribution of the process $A(t)$ be Gaussian:
\begin{equation} \label{PGauss}
P\left[ A \right] = \,\exp \left( { - \frac{1}{4}\int\limits_0^T
{{A_{ij}}D_{ijkp}^{ - 1}{A_{kp}}dt} } \right) \ ,
\end{equation}
where $D_{ijkp}$ is determined by the pair correlation function,
\[2{D_{ijkp}}\delta \left( {t - t'} \right) =
\left\langle {{A_{ij}}\left( t \right){A_{kp}}\left( {t'} \right)}
\right\rangle \] Isotropy requires
$$
{D_{ijkp}} = a{\delta _{ij}}{\delta _{kp}} + b{\delta _{ik}}{\delta
_{jp}} + c{\delta _{ip}}{\delta _{jk}}
$$
Here $a,b,c$ are constants (they must satisfy the condition $P[A]>0$
for any $A$). The cumulant function corresponding to (\ref{PGauss})
is
\begin{equation} \label{wGauss}
w\left( \eta  \right) =   {\eta _{ij}}D_{ijkp}^{}{\eta _{kp}} =
a\,{\left( {tr\eta } \right)^2} + b\,tr\,\eta {\eta ^T}
+c\,tr\,{\eta ^2}
\end{equation}
From (\ref{lamb1}) we  then get
\begin{equation} \label{lambdaGauss}
\lambda _s^G = D\,\left( {2s - 1 - N} \right)  \  , \qquad D=b+c
\end{equation}
We note that the Gaussian spectrum is antisymmetric relative to the
change $s \to N - s + 1$; in particular,
\[\sum\limits_s {\lambda _s^G}  = 0\]

In many applications one needs the additional restriction $tr A=0$.
From $\langle (tr A)^2 \rangle =0$
 it then follows %
\footnote{The matrix $D_{ijkp}^{-1}$ then becomes singular, which
formally corresponds to appearance of the multiplier
$\prod\limits_t {\delta \left( {tr\,A\left( t \right)} \right)} $.}   %
$$
Na+b+c=0
$$
Since $a$ does not contribute to $\lambda_s$, this condition
does not affect the spectrum (\ref{lambdaGauss}).

\vspace{0.7cm}

To describe the statistics of $X_{ss}=\rho_s$ more accurately,  one
can also calculate  its mean-square  deviation from the average:
denote
$$
{\xi _s} = {\rho _s} - {\lambda _s} \ ,
$$
then
\begin{equation} \label{ksiksi}
 \left\langle {{\xi _s}\left( {{t_1}}
\right){\xi _q}\left( {{t_2}} \right)} \right\rangle  = \left\langle
{{\rho _s}\left( {{t_1}} \right){\rho _q} \left( {{t_2}} \right)}
\right\rangle_c =
 \left\langle
{{\rho _s}\left( {{t_1}} \right){\rho _q} \left( {{t_2}} \right)}
\right\rangle  - {\lambda _s}{\lambda _q}
\end{equation}
According to (\ref{Xw}),
\begin{equation} \label{ksi1}
 \left\langle {{\xi
_s}\left( {{t_1}} \right){\xi _q}\left( {{t_2}} \right)}
\right\rangle  = D_{sq} \delta \left( {{t_1} - {t_2}} \right) \ ,
\qquad D_{sq}= \frac{\partial }{{\partial {\eta
_{ss}}}}\frac{\partial }{{\partial {\eta _{qq}}}}{w}\left( \eta_0
\right)
\end{equation}
Simplifying the calculation by using only diagonal components of
$w(\eta)$ in accordance with (\ref{wdiag}), for the Gaussian
function (\ref{wGauss}) we have
$$ D_{sq}^G = 2\left( {D{\delta _{sq}} + a} \right)
$$

Unlike the Lyapunov exponents, the dispersion of $\rho_s$ depends on
the distribution of $tr A$. In the case of traceless Gaussian
matrices we get the well-known relation derived by
\cite{FalkGawVerg}:
\[\left\langle {{\xi _s^G}\left( {{t_1}} \right){\xi _q^G}\left( {{t_2}} \right)} \right\rangle
= 2D\left( {{\delta _{sq}} - \frac{1}{N}} \right)\delta \left(
{{t_1} - {t_2}} \right)\]

\subsection{Non-Gaussian process}

Let now $A(t)$ be a non-Gaussian isotropic $\delta$-process. As we
have seen in the beginning of the section, for diagonal components
of $ \langle X_{sq} \rangle $  one can use (\ref{lamb1}) with
$w(\eta)$   replaced by its diagonal part (\ref{wdiag}):
\begin{equation} \label{lambdawdiag}
{\lambda _s} = \left.  \frac{\partial }{\partial {\eta _{ss}}}
 w_d\left( \eta_{11}\dots\eta_{NN}    \right)  \right|_{\eta=\eta_0}
 \end{equation}

The isotropy condition (\ref{wisotr}) means that $w_d$ must be a
function of combinations $\sum \limits_s \left( \eta_{ss} \right)
^n$ with different $n$:
\begin{equation}  \label{f}
w_d\left( \eta_{11}\dots\eta_{NN}    \right) = f \left(  \sum
\limits_s \left( \eta_{ss} \right) , \sum \limits_s \left( \eta_{ss}
\right) ^2, \sum \limits_s \left( \eta_{ss} \right) ^3,\dots \right)
\end{equation}
Decomposing $w_d$ into symmetric and antisymmetric (relative to
$\eta \to -\eta$) parts, we get
\[{w_d}\left( \eta  \right) =
w_d^ + \left( \eta  \right) + w_d^ - \left( \eta  \right) \ , \qquad
w_d^ {\pm} \left( \eta  \right) = \frac{{w_d^{+}\left( \eta  \right)
\pm w_d^{-}\left( { - \eta } \right)}}{2} \] One can see from
(\ref{Aw}) that the symmetric part of $w_d$ contributes to the
connected even order correlation funcitons of $A$, and $w_d^{-}$
contributes to those of odd orders. Furthermore,  we decompose
\begin{equation} \label{sviaz}
{\lambda _s} = \lambda _s^{-} + \lambda _s^{+} \ ,
\end{equation}
here $\lambda _s^{+}$ is produced by the antisymmetric part of
$w_d$, and vice versa:
\begin{equation} \label{lambsim}
\lambda _s^{\pm} = \left.  \frac{\partial }{{\partial {\eta
_{ss}}}}w_d^{\mp} \right|_{ \eta_0 }
\end{equation}
Then, symmetric properties of $w_d$  cause the symmetric properties
of $\lambda_s$ relative to $\eta_0 \to -\eta_0 $,  i.e., $s \to
N+1-s$:
$$
\lambda^{\pm}_{N+1-s} =\pm \lambda^{\pm}_s
$$
Hence, the symmetric part $\lambda_s^{+}$ is determined by the
odd-order correlators of $A$, and the asymmetric part
$\lambda_s^{-}$
depends on the  even-order correlators.%
\footnote{Note that in the Gaussian case the Lyapunov spectrum is
odd because there are no non-zero odd-order correlators.} %
 This is an important feature
of the random processes: e.g., in theory of turbulence it is
connected with the asymmetry of a turbulent flow relative to the
change of time direction (\cite{ZSpre,IZ}).

If the matrices $A$ (and hence, $X$) are assumed to be traceless, we
have
\[\sum\limits_s {\lambda _s^{+} = } \sum\limits_s {\lambda _s^{}}
 = \sum\limits_s {\left\langle {X_{ss}^{}} \right\rangle }  = 0\]

Generally, the Lyapunov spectrum is determined by the choice of $f$
in (\ref{f}). Here we analyze one particular (important for physical
applications) case of traceless $3\times3$ matrices $A(t)$ with
distribution close to the Gaussian.

For $N=3$ and $tr A=0$, the Lyapunov spectrum takes the form
\begin{equation}   \label{lam3}
\left\{
\begin{array}{l}
{\lambda _1} =  - \Delta 
 - \frac{{{\lambda _2}}}{2},\,\,\\
{\lambda _2}\\
{\lambda _3} = \Delta 
- \frac{{{\lambda _2}}}{2}
\end{array}  \right.
\end{equation}
 According to (\ref{sviaz}),  $\lambda_2$ is determined
by the odd, and $\Delta 
$ by the even part of $w_d$. If we restrict (\ref{f}) to the
first two terms, 
the traceless condition
$$
 \sum \limits_s \frac{\partial }{{\partial {\eta _{ss}}}} w_d =0
$$
 gives
\begin{equation} \label{wdseries}
\begin{array}{rcl}
{w_d}\left( \eta  \right) &=& \textstyle D\left( { \sum\limits_q
{\eta _{qq}^2} - \frac{1}{3}{{\scriptstyle\left( {\textstyle
\sum\limits_q {{\eta _{qq}}} } \scriptstyle\right)}^2}} \right) \\
&+& F\left( {\textstyle\left( { \sum\limits_q {\eta _{qq}^2} }
\right)\left( { \sum\limits_q {{\eta _{qq}}} } \right) -
\sum\limits_q {\eta _{qq}^3}  - \frac{2}{9}{{\left( {\sum\limits_q
{{\eta _{qq}}} } \right)}^3}} \right) \end{array}
\end{equation}
The coefficient $F$ is called the asymmetry coefficient of the
process; in accordance with (\ref{Aw}), it determines the
third-order correlation of $A$, for example
$$
\left\langle {{A_{11}}\left( {{t_1}} \right){A_{11}}\left( {{t_2}}
\right) {A_{11}}\left( {{t_3}} \right)} \right\rangle  = \, -
\frac{4}{3}F\,\delta \left( {{t_1} - {t_2}} \right)\delta \left(
{{t_1} - {t_3}} \right)
$$
 From (\ref{lambdawdiag})    
 it follows
\[{\lambda _s} = 2D\left( {s - 2} \right) +
2F\left( {1 - \frac{3}{2}{{\left( {s - 2} \right)}^2}} \right)\] So,
\begin{equation} \label{cutlam}
\Delta 
= 2D  \  \qquad {\lambda _2} = 2F
\end{equation}
For the second-order correlator (\ref{ksiksi}), (\ref{ksi1}) we
obtain
$$
D_{sq} =   {2D\left( {{\delta _{sq}} - \frac{1}{3}} \right) +
2F\,\left( {s + q - 4} \right)
 \left( {1 - \frac{3}{2}{\delta _{sq}}} \right)}
 $$
The first term in this equation is Gaussian,  the second term
corresponds to the main non-Gaussian contribution.

In non-Gaussian process, there is also the third-order connected
correlator:
\[
 \left\langle {{\xi _s}\left( {{t_1}} \right)
{\xi _q}\left( {{t_2}} \right){\xi _p}\left( {{t_3}} \right)}
\right\rangle  = {\left\langle {{\rho _s}\left( {{t_1}} \right){\rho
_q}\left( {{t_2}} \right){\rho _p} \left( {{t_3}} \right)}
\right\rangle _c} = {F_{sqp}}\, \delta \left( {{t_1} - {t_2}}
\right)\delta \left( {{t_1} - {t_3}} \right)\] where
\[{F_{sqp}} = \frac{\partial }{{\partial {\eta _{ss}}}}\frac{\partial }{{\partial
{\eta _{qq}}}}\frac{\partial }{{\partial {\eta _{pp}}}}{w_d}\left(
{{\eta _0}} \right)\,\] Substituting (\ref{wdseries}) for $w$, we
get
\[{F_{sqp}} = 2F\left( {{\delta _{sq}} + {\delta _{sp}} + {\delta _{qp}} - 3\,{\delta _{pq}}
{\delta _{sq}} - \frac{2}{3}} \right)\  \ (no \ summation) \]

 The finite-polynomial approximation of $w_d$
has a serious defect:  such $w_d$ corresponds to the probability
density that is not positively defined (\cite{Marcinkevich}). But
still it is a useful simplification, since  the rest of the series
does not make fundamental  changes to lower -order  correlators.

There is a simple way to estimate the validity of cutting the third
term in (\ref{wdseries}). It is known that the Lyapunov indices must
be ordered (\cite{Let3,FalkGawVerg}):
\[{\lambda _1} < {\lambda _2} < {\lambda _3}\]
From (\ref{lam3}), (\ref{cutlam}) we then have
\[\left| F \right| < \frac{{2D}}{3}\]
This is a necessary condition for cutting the series
(\ref{wdseries}). In theory of turbulence, $F$ is also required to
be positive to provide the right sign of energy dissipation (which
is associated with the third-order correlator of $A$).

\section{Statistics of the T-exponential}

In the previous section we discussed different moments of the matrix
$X$. Now we proceed to the averages of its exponentials
(\ref{rho-and-x}). As we mentioned above, the matrix $R$  (which is
the $T$-exponential of $\theta$) separates from the other variables
(\ref{thetaR}). The calculation of $z$  (which is, roughly speaking,
the T-exponent of $\zeta$) is simplified by the fact that
$\zeta^N=0$, so all the series are finite. Here we analyze more
accurately the statistics of $d$. According to  (\ref{rho-and-x}),
\[{d_s}\left( t \right) = \exp \left( {\int\limits_0^t {{\rho _s}\left( \tau  \right)\,d\tau } } \right) = \exp \left( {\int\limits_0^t {{X_{ss}}\left( \tau  \right)\,d\tau } } \right)\]
Thus, its moments are
\begin{equation} \label{dsn}
\left\langle {{{\left( {d_s^{}\left( t \right)} \right)}^n}}
\right\rangle  = \left\langle {\exp \left( {n\int\limits_0^t
{{X_{ss}}\left( \tau  \right)d\tau } } \right)} \right\rangle
\end{equation}
These characteristics are important in the applications; in
turbulence, and in particular in the theory of passive scalar
advection, they describe the separation of trajectories of liquid
particles.

Consider the characteristic functional and the cumulant function of
the $X(t)$ process:
\begin{equation} \label{ZX}
Z^{(X)}\left[ \eta(t) \right] = \exp \left(  \int \limits_0^T
w^{(X)} (\eta) dt \right) = \left \langle  \exp \left(  \int
\limits_0^T tr \left( \eta(t) X(t) \right) dt \right) \right \rangle
\end{equation}
According to (\ref{deltapr}), (\ref{Pdelta}), (\ref{Z}) they are
related to those of the process $A(t)$  by
$$
Z^{(X)} \left[ \eta(t) \right] = Z \left[ \eta(t) + \eta_0  \right]
  Z^{-1} \left[  \eta_0  \right] \ , \qquad  w^{(X)}(\eta)  = w(\eta +
\eta_0) - w(\eta_0)
$$
In the previous section we saw that to calculate the moments of $X$,
one can reduce $w(\eta)$ to the 'diagonal' function $w_d \left(
\eta_{11},\dots,\eta_{NN} \right)$ (\ref{wdiag}). Similarly, to
calculate (\ref{dsn}), it is enough to consider
$$
 w^{(X)}_s (\eta_s) \equiv w^{(X)} (\eta)
\left| _{ \eta_{kp}= \delta_{ks} \delta_{ps} \eta_s \, ,  } \right.
\quad no \ summation
$$
Then from (\ref{lamb1}), (\ref{ksi1}), (\ref{Xw}) it follows that
$w^{(X)}_s$ can be expanded into the series:
\[w_s^{(X)}\left( {{\eta _s}} \right) = {\lambda _s}{\eta _s} + \frac{{{D_{ss}}}}{2}\eta _s^2
  + \frac{{{F_{sss}}}}{{3!}}\eta _s^3 + ...\]
where the first coefficient is the Lyapunov index (\ref{lambdadef}),
the second is the covariation matrix, and the next coefficients
correspond to higher correlators of $X$.

From (\ref{dsn}), (\ref{ZX}) we have
\begin{equation} \label{last}
\left\langle {{{\left( {d_s^{}\left( t \right)} \right)}^n}}
\right\rangle  = \exp \left(  w^{(X)}_s (n) t \right) = \exp \left(
\left(  \lambda_s n + \frac{D_{ss}}2 n^2  +
\frac{{{F_{sss}}}}{{3!}}n^3 +  \dots \right) t   \right)
\end{equation}
We see that the moments increase exponentially with time. From
(\ref{last}) it also follows that  in calculation of moments of the
exponentials (\ref{dsn}), one can not neglect the contributions of
higher-order connected correlators even if all the moments of $X$
are dominated by non-connected diagrams.
 I.e., even if the integral $\int X dt$ satisfies the
condition of the central limit theorem and acts as a Gaussian, $\exp
\int X dt$ is still essentially non-Gaussian.

\section{Smoothed variables}

The cumulant function determines completely the statistical
properties of a $\delta$-process. However, it cannot be measured in
experiments (though the statistical averages, which are its
derivatives, are measurable), nor is the probability density
functional a measurable value. (Even more, the PDF is, as opposed to
$w(\eta)$,  badly defined in the case of non-Gaussian
$\delta$-process, it requires renormalization.) In this section, we
introduce smoothed random variables  which can be measured in
experiments. We show that they appear to have same cumulant function
as the $\delta$-process, and their averages are related strongly to
the correlators of $A$ and $X$.

Consider a random matrix
\[\bar A = \int\limits_0^1 {A(t)dt} \]
The corresponding characteristic  function is
\begin{equation} \label{zsmall}
z\left( \eta  \right) = \left\langle {\exp \left( {tr\left( {\eta
\bar A} \right)} \right)} \right\rangle
\end{equation}
It is related to the characteristic functional $Z\left[ \eta(t)
\right]$  and its cumulant function $w(\eta)$ by
\begin{equation} \label{zsmallw}
z\left( \eta \right) = Z\left[ {\eta \theta \left( {1 - t} \right)}
\right] = \exp \left( \int\limits_0^T {w\left( {\eta \theta \left(
{1 - t} \right)} \right)  dt} \right)  = \exp \,w\left( \eta \right)
\end{equation}
 So,  $w(\eta)$ is the
cumulant function for both the stochastic process $A(t)$ and
stochastic variable  ${\bar A}$. The averages
$$
 \left\langle {{{\bar A}_{ij}}...{{\bar A}_{kp}}} \right\rangle
=  \frac{\partial }{{\partial {\eta _{ij}}}}...\frac{\partial }
{{\partial {\eta _{kp}}}} \exp \left( w(0) \right)
$$
 can be obtained from the corresponding correlators
(\ref{Aw}) by omitting the delta-functions, the  terms
$$
 \left\langle {{{\bar A}_{ij}}...{{\bar A}_{kp}}} \right\rangle _c
=  \frac{\partial }{{\partial {\eta _{ij}}}}...\frac{\partial }
{{\partial {\eta _{kp}}}}  w(0)
$$
 are equal to the coefficients  of the connected correlation functions.

One can express $w(\eta)$  by means of the probability density
$p(\bar{A})$ :  from (\ref{zsmall}), (\ref{zsmallw}) one can easily
derive
\[w\left( \eta  \right) = \ln \int {d\bar A\,p\left( {\bar A} \right)} \exp \left( {tr\left( {\eta \bar A} \right)} \right)\]
For example, if $A(t)$ is a Gaussian  process (\ref{PGauss}) then
$p(\bar A)$ is Gaussian, too:
\[p\left( {\bar A} \right) = \,\exp \left( { - \frac{1}{4}{{\bar A}_{ij}}D_{ijkp}^{ - 1}
{{\bar A}_{kp}}} \right)\] and corresponds to the  same cumulant
function (\ref{wGauss}). In the general case, $p(\bar A)$  can be
written in the form
\[p\left( {\bar A} \right) = \,\exp \left( { - \frac{1}{4}{{\bar A}_{ij}}D_{ijkp}^{ - 1}
{{\bar A}_{kp}} - V\left( {\bar A} \right)} \right)\] where the
function $V\left( {\bar A} \right)$ corresponds to the non-Gaussian
part. Then the  cumulant function is
$$
w\left( \eta \right) = \ln \left( {\exp \left( { - V\left(
 {\frac{\partial }{{\partial \eta }}} \right)} \right)\exp
 \left( {{\eta _{ij}}D_{ijkp}^{}{\eta _{kp}}} \right)} \right)
 $$
where $\exp \left( { - V\left( {\frac{\partial }{{\partial \eta }}}
\right)} \right)$ should be understood as a formal series.  This
allows, in particular, to restore the cumulant function by known
coefficients in connected correlators. Now as we have the smoothed
variables,  the cumulant function is not necessary: $p(\bar{A})$
contains all the information about the delta-correlated random
process $A(t)$, and
$$
\left \langle X_{ij}(t_1)...X_{kp}(t_n) \right \rangle _c = N \int d
\bar A \, \bar{A}_{ij}...\bar{A}_{kp} \, p(\bar{A}) \, \exp\left( tr
(\eta_0 \bar{A}) \right) \delta (t_1-t_2) ...\delta (t_1-t_n) \ ,
$$
$$
N=\left(  \int d \bar A  p(\bar{A}) \exp\left( tr (\eta_0 \bar{A})
\right)   \right)^{-1}
$$
In particular, 
$$
\lambda_s = N \int d \bar A \, \bar{A}_{ss} \, p(\bar{A}) \,
\exp\left( tr (\eta_0 \bar{A}) \right)
$$

\section{Conclusion}

In this paper we present a very simple technics to calculate the
averages of time-ordered exponentials of random $N\times N$ matrices
(\ref{Q-Texp}) and, thus, describe the evolution of linear
stochastic systems.

The simplification comes from changing  the variables
(\ref{change}), (\ref{decA}) that allows to separate the rotations
of the eigenvectors  and therefore to 'stabilize' the rest of the
solution. The formal solution (\ref{Q-Texp}) can then be rewritten
as a functional integral (\ref{averageX}) with a simpler
'rotational' T-exponent (\ref{R-Texp}).

If the random process is isotropic, the 'rotational' part can be
excluded, and the averages (\ref{averageX}) become functional
integrals without time-ordered products. In the case of $\delta$-
correlated in time process
 we calculate the
Lyapunov spectrum (\ref{lambdadef}) and other correlation
characteristics of the T-exponent (\ref{Xw}). %

For Gaussian probability distribution of the matrices, the results
coincide with those obtained earlier by \cite{GambaKol}; the
non-Gaussian case is analysed in details. We also describe the
relation of the statistics of the T-exponentials to the statistics
of 'smoothed' variables. They appear to be determined by same
cumulant function, so that, measuring probability density  function
(or correlators) of the 'smoothed' matrices, one can easily
calculate, e.g. the Lyapunov spectrum.

The important feature of the non-Gaussian probability distribution
is that the higher-order connected correlation functions cannot be
neglected when counting the exponential averages, even if the
difference from the Gaussian is small and higher-order correlators
of the exponent are dominated by the Gaussian (non-connected)
contribution. One more difference from Gaussian is that the Lyapunov
spectrum of a non-Gaussian process is, generally, asymmetric. This
is very important for applications. In particular, in \cite{ufnZS}
the observed scaling space distribution of velocity and statistical
properties of a turbulent flow were derived  from the statistics of
velocity deformation tensor at large scales based on a stochastic
analog to Euler equation.

It was shown that symmetry of the Lyapunov spectrum corresponds to
time invariance of the flow, and hence must be broken in real flows.
The right sign of energy flux (from larger to smaller scales in
3$^d$) requires $\langle tr A^3 \rangle <0$. Thus, one cannot
restrict the consideration by Gaussian approximation:
non-Gaussianity is of crucial importance. The results achieved in
this paper allow to calculate the Lyapunov spectrum based on
experimental measurements of large-scale velocity statistics, then
derive the statistic characteristics of velocity field at small
scales in accordance with \cite{ufnZS}, and compare them to the
observations.

This work is supported by the RAS program 'Nonlinear dynamics in
mathematical and physical sciences'.

\end{document}